\documentclass[11pt]{article}
\usepackage[english]{babel}
\usepackage{graphicx}
\usepackage[authoryear,round]{natbib}
\usepackage{rotating}
\usepackage[margin=25mm]{geometry}
\usepackage{hyperref}
\usepackage[font=small]{caption}
\hypersetup{
    colorlinks = true,
    linkcolor = blue,
    anchorcolor = blue,
    citecolor = blue,
    filecolor = blue,
    urlcolor = blue,
	pdfauthor={L. Foschini},
	pdftitle={The Cut of Time}
    }
\usepackage{amssymb,amsmath}
\usepackage{url}
\setcitestyle{aysep={}}

\begin{document}

\title{The cut of time}
\author{Luigi Foschini\\
\small{Brera Astronomical Observatory}\\
\small{National Institute of Astrophysics (INAF)}\\
\small{Milano/Merate (Italy)}\\
\small{email: \texttt{luigi.foschini@inaf.it}}
}
\date{December 1, 2026}
\maketitle

\begin{abstract}
After a short review of how time is treated, or neglected, in classical physics, quantum mechanics, relativity, and quantum gravity, I suggest changing the interpretation of time, from a duration to a cut. The now is the cut dividing a past that no longer exists from a future that does not exist yet and must be generated. Time reversibility turns out to be an idealisation, broken by energy dissipation and, at the microscopic level, by the weak interactions. In quantum mechanics, the cut separates the before and the after of a measurement, and the correlations of entangled particles are born together with the particles themselves, without denying non-locality in Bell's sense. In relativity and quantum gravity, the cut can be identified with the lapse function of a privileged cosmological foliation, and represented through the imaginary exponential that describes the shift in time of a quantum state. This view finds a natural counterpart in recent proposals in which space-time remains classical while matter is quantum, and in the idea that time is more fundamental than space. A reassessment of terminology is also required to avoid meaning traps, from the ambiguity of the word \emph{future} to the misleading expression \emph{time reversal violation}. Finally, I address the problem of estimating the energy needed by a physical object to shift in time without any other interaction or motion, that is, only to exist. Such an existence energy would imply a tiny violation of the invariance under time translation: I discuss how it can be constrained by the lifetimes of protons and electrons, and I estimate the energy associated with the expansion of space by an elementary amount.
\end{abstract}

\hangindent=8cm
\hangafter=0
\begin{flushright}
\emph{``What is time?'' -- the error is already contained in the question, as if the question were: of what, of what material, is time made? As one might say -- of what is this fine dress made?}\\Ludwig Wittgenstein (\citeyear{WITTGENSTEIN})
\end{flushright}
\normalsize

\section{Introduction}
Time is intrinsically different from space, but it has been often spatialized or, more often, neglected. The difference should be apparent at the very moment one wants to measure it. If one measures the size of an object, he takes a ruler, puts it on one side of the object, and reads the numbers on it. It is not possible to measure time in this way. There is no ruler between now and then, yesterday and today. One can only make the hypothesis that time flows with constant and uniform speed, so to use cyclic motion as ruler for time. A pendulum, a mechanical clock, the heart rate, repeating one word or one number (e.g. 1001, 1002, 1003,...), an atomic clock... In addition, the cyclic motion also requires the assumption that the state at the end of each cycle is exactly the same as the state at its beginning, and that the laws governing the device do not change from one cycle to the next (homogeneity of time, or invariance under time translation). Strictly speaking, this is never exactly true. If nothing else, because of the cosmological expansion, the point of return lies in a universe slightly different from that of the starting point: for each second of the cycle, the scale factor of the universe grows by a fraction $H_0\,\Delta t \approx 2\times 10^{-18}$, and the temperature of the cosmic microwave background decreases by the same fraction. Local bound systems, such as a pendulum or an atom, do not expand, but they are immersed in a universe that does. The difference is negligible in practice, but it is not zero, and it should not be neglected when reasoning down to the Planck scale (see also Sect.~\ref{sec:exist}). The assumption of identical cycles must not be confused with time reversibility (invariance under time reversal): the two are often mixed up, but they are distinct, and both will be relevant in the following. Paradoxically, modern metrology has reversed the relationship between space and time: since 1983, the metre is defined through the second and the speed of light, so that lengths are actually measured via time.

The hypothesis of a universal time, flowing at the same rate for everyone, was never verified. On the contrary, the development of devices based on relativity (e.g. Global Positioning System, GPS) clearly showed that there is no universal rate of time: the rate of a clock depends on its motion and on the gravitational field, although each clock, locally, ticks according to the same physical laws. As known from almost all textbooks on this topic (e.g. \citealt{CHENG}; for a detailed treatment, see \citealt{ASHBY}), according to special relativity, satellite clocks are slower than those on ground by $\sim 7$~$\mu$s per day, because the former have an orbital speed of $\sim 4$~km/s. Also general relativity implies a different rate: as satellites are at an altitude of $\sim 20000$~km, the weaker gravitational field makes clocks run faster by $\sim 46$~$\mu$s per day. The net effect is that satellite clocks gain $\sim 38$~$\mu$s per day ($\sim 19$~$\mu$s per orbit), which must be corrected. As one microsecond is equal to about three hundred metres on the ground, such a seemingly small drift would translate, if uncorrected, into an error on the ground of $\sim 11$~km per day.

The hypothesis of an absolute time was born with classical physics, where dealing with macroscopic objects implied negligible perturbations, making it possible to neglect time or to consider it as a space dimension. Later, relativity and quantum mechanics made it impossible to neglect time. However, the spatialization remained, the reference to the duration, particularly in relativity, where time is often considered as the fourth dimension. The criticism of the spatialization of time goes back at least to \citet{BERGSON}, who proposed to replace it with the duration (\emph{dur\'ee}); in this essay, I propose the opposite remedy. Aim of the present essay is to emphasise the role of time in physics and to propose a different interpretation. After a short review on how time is taken, or not taken, into account in some branches of physics (more examples and details can be found in \citealt{FOSCHINI}), I will show that thinking of time as a cut could open interesting questions and remove some paradoxes. It is worth noting that I am just proposing to recover the original meaning of time. The Italian word \emph{tempo} derives from Latin \emph{tempus}, which, according to a traditional etymology, is linked to the Greek \emph{tem-n\^ o}: to cut, to divide (see, however, \citealt{DEVAAN} for a different interpretation)\footnote{Interestingly, the same Greek verb is at the origin of the word \emph{atom}, from \emph{a-tomos}, that which cannot be cut: the atom, the basic constituent of matter, would then be something not subject to the cut of time \citep{NOMASS}.}. The English word \emph{time} has a different etymology, coming from the Germanic languages, but is anyway linked to the proto-indo-european root \emph{da-}, which means again to divide\footnote{See \url{http://www.etimo.it/} for Italian words and from \url{https://www.etymonline.com/} for English words.}. Therefore, time as a cut can be seen as the recovery of an ancient meaning of the word. 

It is also necessary another revision of terms. What can happen is generally indicated as the future. This word derives from the Latin word \emph{futurus}, which is the future active participle of the verb \emph{esse}, to be, formed on the stem \emph{fu-}, the same stem as the perfect \emph{fu\=\i} \citep{DEVAAN}. Therefore, in classical Latin, it means something that will {be}. This is a problem, as the verb to be is not the proper verb to indicate something that has still to be generated. Inevitably, it carries meaning of existence. However, the stem \emph{fu-} derives from the Proto-Indo-European root \emph{*bh\=u-}, meaning to grow, to become (e.g. \citealt{MALLORY}; the same root as the Latin \emph{f\=\i o}, to become, and the Greek \emph{phy\^o}, to generate, and \emph{physis}, nature). Therefore, the term is ambiguous: at its root, it carries the meaning of becoming, while in its classical and modern usage the word future has the meaning of something already existing, but not yet reached by the now (it has also the significant drawback of superstition and fate). In this sense, it would be the proper word for a spacetime without matter (see Sect.~\ref{sec:relphys}), for a temporama. But, if one thinks that what can happen is still to be generated, then the other sense, that of becoming, is required. In Italian, there is the proper word, \emph{avvenire}, but it has no correspondence in English. It can be translated with world to come or with forthcoming, although the latter has a meaning of something to happen very close in time; the philosophical literature also uses the expression \emph{open future}. In the following, I will use future, but in the sense of \emph{avvenire}: what is still to be generated, not what already exists and is waiting to be reached\footnote{In a previous version of this essay, \emph{futurus} was described as the future perfect of an archaic form of the verb to be, and \emph{future} was replaced by \emph{forthcoming}. The analysis given here corrects that description and shows that the word is ambiguous, rather than simply inappropriate. Therefore, \emph{future} has been restored, with the caveat discussed in the text.}. What happened has fewer problems of translation: past is still a useful term.

Etymology is not an argument, of course, but it shows how speakers have conceptualised time.

\section{Classical physics}\label{sec:classical}
It is often written and said that basic physics laws are not affected by the direction of time\footnote{See, for example, Sean Carroll in an interview by K. Becker at \url{https://fqxi.org/community/articles/display/236} (2021) or \cite{BARBOUR}.}. As it is possible to move in space toward left or right, up or down, it should be possible to move in time forward and backward (reversibility). Laws of physics do not change if one substitutes $t$ with $-t$. This is often taken as a sign that time is spatialised, as if it were nothing else than one more spatial dimension. However, as already noted by \citet{WITTGENSTEIN}, one could walk along one direction and, then, walk backward exactly on his own footsteps, but this does not mean to go backward in time. Therefore, the simple substitution of $t$ with $-t$ is not equivalent to a spatial inversion ($x\rightarrow -x$). It is also worth keeping distinct two properties that are often confused: invariance under time reversal ($t\rightarrow -t$) and invariance under time translation ($t\rightarrow t+t_0$). The latter states that the laws of physics do not change with time and, by Noether's theorem, it is equivalent to the conservation of energy. This section deals with the former.

A classical example is the linear harmonic oscillator, whose general solution is $x(t)=A\sin \omega t + B \cos \omega t$. This is a periodic oscillation independent of the sign of time: it is the same if one substitutes $t$ with $-t$. This equation is also the basis for the isochronous pendulum. However, this is just an ideal case. If one adds friction, this extra term means that the general solution must be multiplied by an exponential $e^{-\rho t}$ ($\rho$ being a friction term), with a slightly reduced frequency: the result is a damped oscillation and a clear direction of time. If one wants to keep constant the motion, then it is necessary to add energy to the device (old mechanical clocks needed to be wound up each day). If one wants to invert time, then it is also necessary to reverse the energy dissipation due to the friction. Damped oscillations should become amplified, fed by the heat previously dissipated in the environment, which should flow back into the oscillator. Energy would be conserved also in this reversed process: what prevents it from being observed is not the first, but the second principle of thermodynamics, since the reversed process would imply a decrease of entropy.

The same concept can be applied to many particles (thermodynamics): whatever is the starting distribution of particles, after a while it becomes a Maxwell-Boltzmann distribution and the entropy increases (in absence of external energy input). In the ideal case, the time direction due to the increase of entropy is a purely probabilistic matter and, as a matter of principle, it might be reversed, although the expected time for a spontaneous inversion, named after Poincar\'e, is quite large. For an Avogadro number of particles, the Poincar\'e time has been estimated to be $10^{10^{23}}$~s \citep{HUANG}. However, again, it is worth reminding that the above condition is valid in an ideal case (for example, the container of the gas must have rigid walls, particles must be hard spheres,...). The addition of reality elements, i.e. energy dissipation, enforces the direction of time. It must be noted, however, that in classical mechanics friction and dissipation are a phenomenological description of the interaction with a huge number of microscopic degrees of freedom, whose dynamics is time-reversible (this is the core of Loschmidt's objection to Boltzmann). Therefore, within classical physics, dissipation \emph{is} the arrow of time, rather than its cause. As we shall see, the premise of Loschmidt's objection does not hold in general: in quantum mechanics, the measurement introduces an irreversible cut (Sect.~\ref{sec:qm}), and even the microscopic laws of weak interactions are not exactly symmetric under time reversal (see below). The deeper question, why the universe started in a state of low entropy (the so-called \emph{past hypothesis}; \citealt{ALBERT}), remains open and is naturally linked with the cosmological time discussed in Sect.~\ref{sec:qg}.

There are also other reasons to doubt that classical mechanics is really time-reversible and deterministic. There are cases where the solutions of Newton's equations are not unique \citep{HUTCHISON,NORTON}, and it has been argued that the finite information content of physical quantities makes classical physics intrinsically indeterministic \citep{DELSANTO}. Moreover, even at the microscopic level, the laws of weak interactions are not symmetric under time reversal: a process and its time-reversed counterpart, that is the process with the initial and final states exchanged, do not occur with the same probability \citep{BABAR}\footnote{In the language of particle physics, this is called \emph{time reversal violation}, an expression that can be misleading: what is violated is the symmetry of the laws under time reversal, not the direction of time. Nothing goes backward in time: both processes run forward, and they differ only in the order of the initial and final states. The identification of the asymmetry measured by \citet{BABAR} with a violation of time-reversal symmetry requires some assumptions, which are discussed by \citet{APPLEBAUM}.}. This asymmetry is probably unrelated to the thermodynamic arrow of time: it concerns only particular processes of weak interactions, which play no role in the dynamics of an ordinary gas, governed by electromagnetic interactions that are symmetric under time reversal to a very high precision. Moreover, the reversed process is still allowed if time reversal is combined with the exchange of particles and antiparticles and with spatial inversion (CPT symmetry), so that a version of Loschmidt's objection survives. Nevertheless, it shows that nature distinguishes the two directions of time already in its fundamental laws.

To summarise, in classical physics, time reversibility is an idealisation. As soon as one considers energy dissipation, a one-way direction of time becomes evident, although its ultimate origin may have to be searched for in the initial conditions of the universe.

\section{Quantum physics}\label{sec:qm}
In classical physics, the state of a particle is defined when its position $x(t)$ and momentum $p(t)$ are known at a certain time. These quantities should be measured at the same time, which in practice is never exactly possible, although nothing forbids it in principle in classical physics. Again, we are facing an ideal case. It is taken for granted that the measurements made in two consecutive close times, either $x(t)$ and $p(t+\delta t)$, or $x(t+\delta t)$ and $p(t)$, could be considered as done at the same time, because the perturbations on the measured quantities are negligible. Time can be neglected. Therefore, $x(t)\approx x(t + \delta t)$ and $p(t) \approx p(t+\delta t)$, so that it is possible to define the state of a particle via the conjugated variables $x(t),p(t)$.

As known, this is no more possible in quantum mechanics, because of the principle of indeterminacy. Perturbations are no more negligible, which means that $x(t)\neq x(t + \delta t)$ and $p(t) \neq p(t+\delta t)$. Quantum mechanics forces us to take time into account. There is a \emph{before} and an \emph{after} the measurement. This also means that a measurement changes the state of the system, so that incompatible quantities cannot be measured on the same system without disturbing each other (an ideal measurement of the same quantity, repeated immediately, gives instead the same result). One can prepare an ensemble of particles all in the same way, make many measurements and calculate statistics. However, there will always be an intrinsic dispersion of the outcomes, which is a property of the quantum state and not a measurement error: repeating the measurements improves the estimate of the mean value, as in classical physics, but cannot reduce the quantum dispersion itself.

Before the measurement, one does not know where a particle is, but it is possible to estimate the probability. After the measurement, one knows where the particle \emph{was}, as the measurement changed the momentum. The hinge of measurement is not the observer or its interaction, but the irreversibility \citep{BOHR,DANERI}, an idea that today is developed in the framework of decoherence \citep{ZUREK}, although decoherence alone does not explain why a single outcome is obtained. The well-known metaphor of the \emph{collapse of the wave function} is useful to emphasise the key role of the time operating a cut, dividing before and after the measurement. A beautiful example of the time cut is given by the theory of $\beta-$decay by Enrico Fermi \citeyearpar{FERMI}. Before, there is one neutron. After, there are one proton, one electron, and one antineutrino. It is worth noting that the neutron does not \emph{contain} the proton, the electron and the antineutrino. These three particles are generated when the neutron vanishes. In today's Standard Model, one $d$ quark of the neutron becomes a $u$ quark, while the other two quarks persist; nevertheless, the proton as such, the electron, and the antineutrino did not exist before the decay. Strictly speaking, quantum mechanics describes the decay as a continuous evolution of a superposition of the undecayed and decayed states, and the sharp cut appears with the irreversible detection of the products: the $\beta$-decay is then a further example of the cut associated with measurement.

The same reading removes the paradox of Schr\"odinger's cat \citep{SCHRODINGER}. A cat is closed in a box together with a radioactive atom and a device that breaks a vial of poison if the atom decays. According to the superposition principle, before the box is opened the whole system is described by a superposition of a decayed atom with a dead cat and of an undecayed atom with a living cat, so that the cat seems to be both alive and dead. This is a paradox only if time is neglected, that is, if the superposition, which describes what can happen before the measurement, is mistaken for the result of the measurement, which is what happened after it. As already noted by \citet[Chap.~I, Sect.~4]{DIRAC}, the intermediate character of a state formed by superposition shows itself in the probabilities of the possible results, and not in the result itself: an observation never gives a result intermediate between those expected for the original states. The superposition distributes the probabilities between \emph{alive} and \emph{dead} before the cut; after the cut, there is only one cat, either alive or dead. Moreover, in the reading proposed here, the cut does not wait for an observer to open the box: it occurs with the irreversible recording of the decay by the detector, and since then the cat is either alive or dead.

Another important example is entanglement. There is one particle A with spin 0 that decays into two particles, B and C, with opposite spins ($+1/2,-1/2$). Thinking in terms of time cut: the particle A, at the time $t$, generates two particles, B and C, which did not exist before the decay: they were generated when A vanished. Their correlation is born with them, and it is imposed by the conservation of angular momentum: at the cut, A does not generate two objects carrying their own pre-written properties, but a single joint state of the pair (the singlet state), whose total angular momentum is fixed by the conservation law. Therefore, the correlations do not require a mysterious action at a distance between objects that already existed: they are a property of the way in which the pair has been generated. This is not a theory of local hidden variables. Bell's theorem \citep{BELL}, confirmed by loophole-free experiments \citep{HENSEN}, excludes that the observed correlations, particularly when the orientation of the measuring devices is changed, can be explained by local properties determined in advance. Bell tests are statistical, as they are performed on large ensembles of pairs, but their conclusion applies to any description of the single pair. The cut of time does not claim otherwise: it does not aim to eliminate non-locality in Bell's sense, nor to replace the quantum formalism. What Bell's theorem does not affect are the conservation laws of energy, momentum, and angular momentum, which hold both for the single process and statistically for the whole ensemble. The cut only shifts the ontological attention: correlations emerge at the moment in which the particles are generated and bound by conservation laws. No property at a certain time $t$ is determined or affected by events that could happen after $t$ (if one eats a cake today, this does not affect his body and his life before that meal): the \emph{after} does not determine retroactively the \emph{before}, and the past does not persist as a repository of properties already written.

How this generation articulates with the full statistics of quantum mechanics (measurements in different bases, contextuality) is still an open problem: on this point, the present proposal is programmatic, not conclusive. One possible route is offered by the privileged cosmological time discussed in Sect.~\ref{sec:qg}, which provides an objective ordering of the measurements performed on B and C, even when they are separated by a spacelike interval. There is also a philosophical assonance with the relational interpretation of quantum mechanics by \citet{RQM}, in the refusal of pre-existing absolute properties. However, the two frameworks do not overlap, and the cut of time is not a variant of relational quantum mechanics: the latter relativizes the state of a system with respect to other systems, while the cut of time denies that properties exist before their generation. They are two different ways of saying that not everything is already written. At present, the cut of time is mainly a different ontological reading of what the quantum formalism already describes: it does not replace the quantum rules, nor does it provide an alternative mechanism explaining the correlations in place of the formalism.

\section{Relativistic physics}
\label{sec:relphys}
As known, in classical physics, space and time are separated. The latter is something external, and independent of space. One point in space is a geometric point and it can move as the time flows. In relativistic physics, space and time are merged into spacetime. One point is an event. The measurement is defined as a procedure performed simultaneously with a clock. Again, the key issue is simultaneity. Relativity forces us to take time into account on another side with respect to quantum physics, the side of the large scale.

As the speed of light in vacuum is constant ($c\sim 3\times 10^{5}$~km/s), and given the time employed by the human brain to process information from senses ($\sim 0.2-0.5$~s), each event happening within $(6-15)\times 10^4$~km from us, is considered as instantaneous. A delay can be already perceived in communications with astronauts on the Moon (average distance $\sim 3.8\times 10^5$~km equivalent to $\sim 1.3$ light seconds). Sunlight needs $\sim 8$ minutes to reach the Earth, and the closest star (Proxima Centauri) is $\sim 4.24$ light years distant from us. These examples suggest that time might be considered as the fourth dimension, equivalent to one space dimension. This is the basis of the so-called \emph{block universe}, where past, present, and future exist all together \citep{PUTNAM}: the argument relies on the relativity of simultaneity in a spacetime without matter, and it is usually extended to any solution of Einstein's equations. As there is a panorama for space, there is also a \emph{temporama} for time. Simultaneity has no more absolute meaning: for events separated by a spacelike interval, what is in the past for one observer could be in the future of another (the order of causally connected events, instead, is the same for all observers). Time is then an illusion.

However, this is again an ideal case, because matter exists and curves spacetime. Even with matter, the solutions of Einstein's equations are four-dimensional blocks, but a homogeneous and isotropic distribution of matter, such as that of our universe on large scales, selects a natural family of observers (those at rest with respect to the cosmic microwave background) and, therefore, a privileged cosmic time. This is a property of the solution, not of the laws, but it is a physical property: it gives an objective meaning to a cosmological now (see Sect.~\ref{sec:qg}). Moreover, matter alters the rate of clocks: the closer the clock to a mass, the slower the pace. 

In the ideal case, the equations of gravity are symmetric under time reversal: the time reverse of a falling rock is a rock thrown upward, which is a perfectly valid solution. In the real case, however, the rock is slowed down by the air and stops with an impact, dissipating its energy, so that the process is irreversible, exactly as for the damped oscillator of Sect.~\ref{sec:classical}. Gravity also provides its own form of dissipation, without the need of friction: a binary system emits gravitational waves, loses energy, and its orbit shrinks, as observed first in the binary pulsar PSR B1913+16 \citep{TAYLOR} and then directly in the merger of two black holes \citep{LIGO}. The reverse process, a binary system gaining energy from gravitational waves converging onto it, is never observed. In the same way, black holes are observed, while their time-reversed counterparts, white holes, are not. In all these cases, the equations admit both solutions, and the asymmetry arises from the conditions selected by nature, as for the increase of entropy.

The case studied by Kurt G\"odel deserves particular attention. He developed a solution of Einstein's equations in the case of a non-expanding, but rotating, universe, where closed timelike curves were present \citep{GODEL}. G\"odel was well aware that his model did not describe our universe: being static, it yields no redshift for distant galaxies, as he wrote himself, both at the end of that paper and in his contribution to the volume in honour of Einstein's seventieth birthday \citep{GODEL3}; see also the review by \citet{ELLIS1996}. His point was philosophical: he was interested in a timeless Parmenidean world, and he was primarily concerned with the non-objectivity of the present, rather than with closed timelike curves \citep{ELLIS1996}. He argued that, if a universe compatible with the laws of relativity has no objective lapse of time, then such a lapse is not guaranteed by the laws even in our universe. In the same essay, he added that there also exist expanding and rotating solutions, in which an absolute time might fail to exist as well, and that it is not impossible that our world is a universe of this kind \citep{GODEL3}. He studied these solutions in a later work \citep{GODEL2}, showing that some of them admit a world time, and hence have no closed timelike curves, while others do not; he remained interested in the possible rotation of the universe \citep{YOURGRAU}. Today, a global rotation of the universe is severely constrained by the isotropy of the cosmic microwave background \citep{SAADEH}, as discussed in \citet{NOMASS}. From the point of view of this essay, G\"odel's work shows that a preferential cosmic time, and hence the difference between time and spatial dimensions, is a property of the specific solution describing our universe. If matter does not rotate, the simultaneity of the observers moving with matter can be extended to the whole universe, defining hypersurfaces orthogonal to their motion; if matter rotates, this is no longer possible, and a world time, if it exists, does not coincide with the simultaneity of these observers \citep{ELLIS1996}.

\section{Quantum Gravity}\label{sec:qg}
\begin{figure}[!t]
\begin{center}
\includegraphics[width=0.95\textwidth]{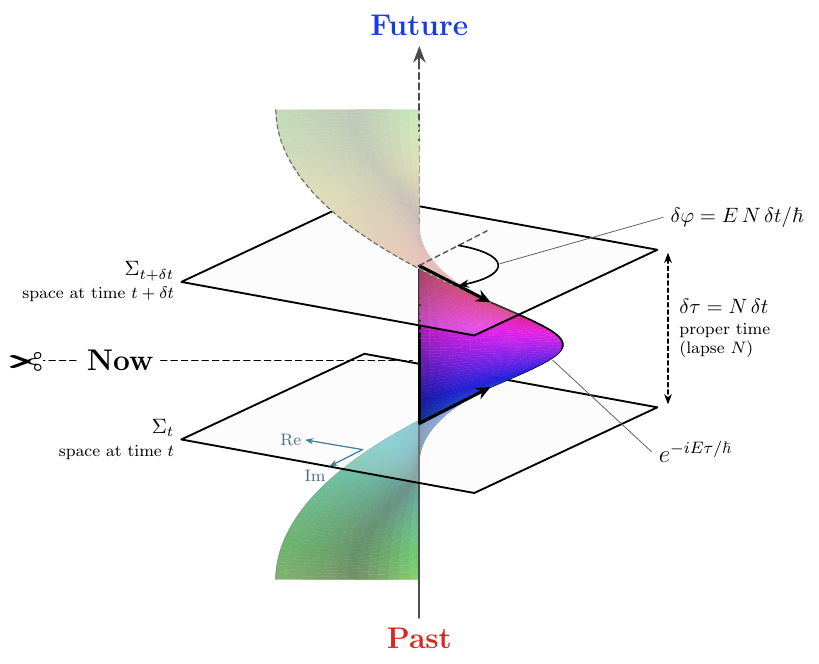}
\caption[$3+1$ spacetime, the imaginary exponential, and the cut of time]{$3+1$ spacetime \citep{ADM}, the imaginary exponential, and the cut of time. The vertical axis is time, and it is also the world line of a particle at rest. The two planes are the hypersurfaces $\Sigma_t$ and $\Sigma_{t+\delta t}$; in this representation, their two directions also serve as the real and imaginary axes of the phase factor $e^{-iE\tau/\hbar}$. The coloured helicoid is the phase, coloured according to its angle: each hypersurface cuts it along one direction (black arrows), and between the two hypersurfaces the phase advances by $\delta\varphi = E\,N\,\delta t/\hbar$, where $\delta\tau = N\,\delta t$ is the proper time set by the lapse function $N$. The now is the cut between the two hypersurfaces: the past (faded) no longer exists, the future (dashed) is still to be generated.}
\label{fig1}
\end{center}
\end{figure}

There are many theories about quantum gravity. I think that the most promising approach is to start from a $3+1$ spacetime, organised as three dimensional hypersurfaces with constant time \citep{ADM,KUCHAR,ELLIS,ELLIS2}. This fits well the cosmological spacetime, but can work well also on smaller scales, for example in the study of black holes \citep{TPM}. The lapse function sets up the relationship between the proper time and the coordinate time: the proper time elapsed between two hypersurfaces, measured along their normal, is the product of the lapse and of the interval of coordinate time (Fig.~\ref{fig1}). However, in general relativity the lapse function expresses a gauge freedom, i.e. any foliation is allowed, and when the theory is quantized, time disappears from the fundamental equation (Wheeler--DeWitt): this is the well-known \emph{problem of time} \citep{KUCHAR2,ISHAM}. Therefore, to interpret the lapse as the cut of time, it is necessary to assume that one foliation is physically privileged. This is not an \emph{ad hoc} assumption: in some approaches, a privileged foliation is built into the theory, such as the constant mean curvature slicing (``York time'') of shape dynamics \citep{SHAPE}, Ho\v{r}ava--Lifshitz gravity \citep{HORAVA}, and causal dynamical triangulations \citep{CDT}; causal set theory, with its models of sequential growth, even describes spacetime as something that is continuously generated \citep{SORKIN}. I suggest interpreting the lapse of the privileged foliation as the cut of time, dividing what happened from what can happen. It is worth noting that the lapse depends on the position on the hypersurface: the cut is global as an ordering, but the proper time elapsed between two hypersurfaces is local, as shown by the GPS example.

It is also worth considering the possibility that gravity should not be quantized at all. \citet{OPPENHEIM2023} has proposed a theory in which space-time remains classical, while matter is quantum, and the two are coupled through a stochastic dynamics; the arguments usually given in favour of the quantization of gravity are reviewed, and found not compelling, in \citet{OPPENHEIMESSAY}. In this view, quantum theory may simply require a classical flow of time, which is very close to the cut of time proposed here. Moreover, the stochastic nature of the coupling implies a fundamental breakdown of predictability, in agreement with the idea that the future is not already written and must be generated (Sect.~\ref{sec:cut}), and it has been argued that the localisation of quantum systems would be induced by their coupling with classical space-time, without the need of a separate measurement postulate (cf. the cut associated with measurement, Sect.~\ref{sec:qm}). The theory can be tested. It predicts a trade-off between the decoherence of quantum systems and the diffusion of space-time, which is already constrained by precision measurements \citep{OPPENHEIMNC}. Moreover, a classical gravitational field cannot generate entanglement, so that the observation of entanglement induced by gravity between two masses \citep{BOSE,MARLETTO} would rule it out.

The cut of time can also be represented by means of the imaginary exponential, which in quantum mechanics describes the shift in time of a physical state (Fig.~\ref{fig1}). The passage from the hypersurface at time $t$ to that at time $t+\delta t$ multiplies the state by the factor $e^{-iE\,\delta\tau/\hbar}$, where $E$ is the energy and $\delta\tau = N\,\delta t$ is the proper time set by the lapse function. The shift in time is then a rotation of the phase in the complex plane, whose rate is set by the energy: each hypersurface cuts the helicoid of the phases along one direction, and between two hypersurfaces the phase advances by $\delta\varphi = E\,N\,\delta t/\hbar$. For a particle at rest, $E=mc^2$, and the phase rotates at the Compton frequency $mc^2/h$, about $1.2\times 10^{20}$~Hz for the electron: this is the same rest energy that is discussed in Sect.~\ref{sec:exist} as the energy associated with the pure shift in time. An overall phase is not observable by itself, but phase differences between different paths are, as shown by neutron interferometry in the gravitational field of the Earth \citep{COW}.

\begin{figure}[!ht]
\begin{center}
\includegraphics[width=\textwidth]{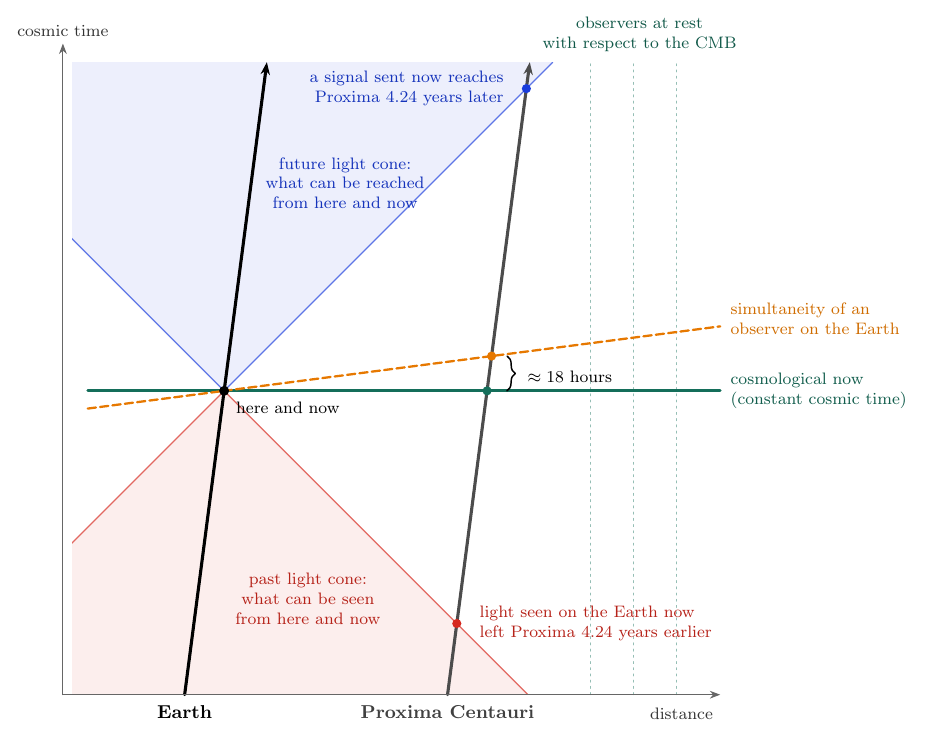}
\caption[The cosmological now and the simultaneity of an observer on the Earth]{The cosmological now and the simultaneity of an observer on the Earth, in a space-time diagram with a single spatial dimension, along the direction of Proxima Centauri. Observers at rest with respect to the cosmic microwave background (CMB) move along vertical lines, and the cosmological now is a hypersurface of constant cosmic time (horizontal line). The Earth and Proxima Centauri move with respect to the CMB, so that the simultaneity defined by an observer on the Earth through the exchange of light signals (dashed line) is tilted: at the distance of Proxima Centauri, it differs from the cosmological now by about 18 hours. The light cones show that the light seen on the Earth now left Proxima Centauri 4.24 years earlier, and that a signal sent now will reach it 4.24 years later. The tilts of the world lines and of the simultaneity line are exaggerated by a factor of about 260.}
\label{fig:now}
\end{center}
\end{figure}

The cosmological time sets up again the simultaneity, although in a different way. A hypersurface corresponds to the space of a cosmological now, which is valid for each point on the surface. On the Earth, one cannot know in real time what is happening on Proxima Centauri \emph{now}, because of the constant and limited speed of light. Nevertheless, this does not mean that nothing is happening. The cosmological now can also be identified in practice: the Earth moves at about 370~km/s with respect to the cosmic microwave background, so that, at the distance of Proxima Centauri, the conventional simultaneity of an observer on the Earth differs from the cosmological now by about 18 hours (Fig.~\ref{fig:now}). The simultaneity becomes an abstract, but useful, idea, like the rigid body or the material point in classical physics.

\section{Time as cut}\label{sec:cut}
The language of classical physics, including the measurements, neglected the time. This was necessary, as many other abstractions, to find universal laws and to build the bases of a reliable science. However, today it is necessary to remind what was neglected. It is not true that the laws used to describe real systems are invariant under time reversal. It is rather obvious that if one removes time or its effects (dissipation of energy), then equations are invariant under time reversal! But this is an ideal case, not the reality. As soon as dissipation is included, equations are no more invariant under time reversal, and even at the microscopic level the laws of weak interactions are not symmetric under time reversal (\citealt{BABAR}; see Sect.~\ref{sec:classical}).

Time, neglected in classical physics, emerged overwhelmingly in quantum mechanics and relativity, although in different ways, and it exploded when trying to elaborate a theory of quantum gravity. In quantum mechanics, time is essential to cut before and after a measurement. One can call it collapse of the wave function or invent another name, but what is important is that time cannot be neglected anymore. Bohr's complementarity can be read in the same way: the cut of time keeps consistency between mutually exclusive behaviours (wave/particle), revealed by mutually exclusive experimental arrangements. Relativity has shown that there is no universal rate of time: the rate of clocks depends on motion and on the gravitational field, which in turn is generated by energy and momentum. The large-scale homogeneity and isotropy of the expanding universe set up a preferential time, making it something different from spatial dimensions. Therefore, the best approach to a theory of quantum gravity is a $3+1$ spacetime, where the lapse of a privileged foliation is reinterpreted as the cut dividing constant-time hypersurfaces.

Being a cut, there is no more a minimum time, it is useless to ask what is its size. It is like asking what is the minimum size of a line one could draw: it depends on the pencil. This does not hamper the concept of interval, but one should not think of the duration. It is not an interval of integers, rather an interval of real numbers. Like real numbers, there is an order, but there is no minimum quantity, no quantum: between any two instants, as between any two real numbers, there are infinitely many others, and no instant has a \emph{next} one. In this sense, the now is like a Dedekind cut, which divides the real line into two parts. It must be noted, however, that the absence of a minimum time is the most natural reading of the cut, but not a logical consequence of it: the continuity of time is a hypothesis on nature. A cut can also be defined in a discrete ordered set, as the boundary between the elements that already exist and those that do not exist yet. This is what happens in the models of sequential growth of causal sets \citep{SORKIN}, where time is discrete, but the universe grows by the addition of new elements. Moreover, the question whether time is discrete is logically independent of the question whether gravity should be quantized: the quantization of gravity does not necessarily imply a quantum of time, and a classical space-time, as in the theory of \citet{OPPENHEIM2023}, is naturally described as a continuum.

One could still invoke the Planck scale, but it could be a transition from a spacetime language to a time-only one, as suggested by \citet{MARKOPOULOU}. According to Markopoulou, the timelessness of general relativity concerns only the geometric time, while a fundamental time exists: what does not fundamentally exist is space, geometry, and gravity. A similar view, with time and energy as primary quantities and space emerging from the relationships between particles, is discussed in \citet{NOMASS}. In the model of quantum graphity \citep{QG2006,QG2008}, the early universe is in a high-temperature phase without locality, that is without space, and geometry emerges only after a phase transition, named geometrogenesis. This distinction between a fundamental time and a geometric time overlaps with the one proposed here, between the cut of time, which is always present, and the proper time measured by clocks, which emerges with space and depends on the observer.

In classical general relativity, however, time does not survive the disappearance of space. The initial singularity of the standard cosmological models is spacelike: it is an instant, lying in the past of every observer, and it is not a point of space-time, but its boundary. Therefore, it makes no sense to ask what happened before. To have time without space, one has to go beyond general relativity, as in the pre-geometric phase of quantum graphity. It is also worth noting that the finite age of the universe refers to the proper time of comoving observers: by using other time variables, such as the logarithm of the volume of space, the initial singularity is pushed to an infinitely remote past \citep{MISNER}.

\begin{figure}[tbp]
\begin{center}
\includegraphics[width=\textwidth]{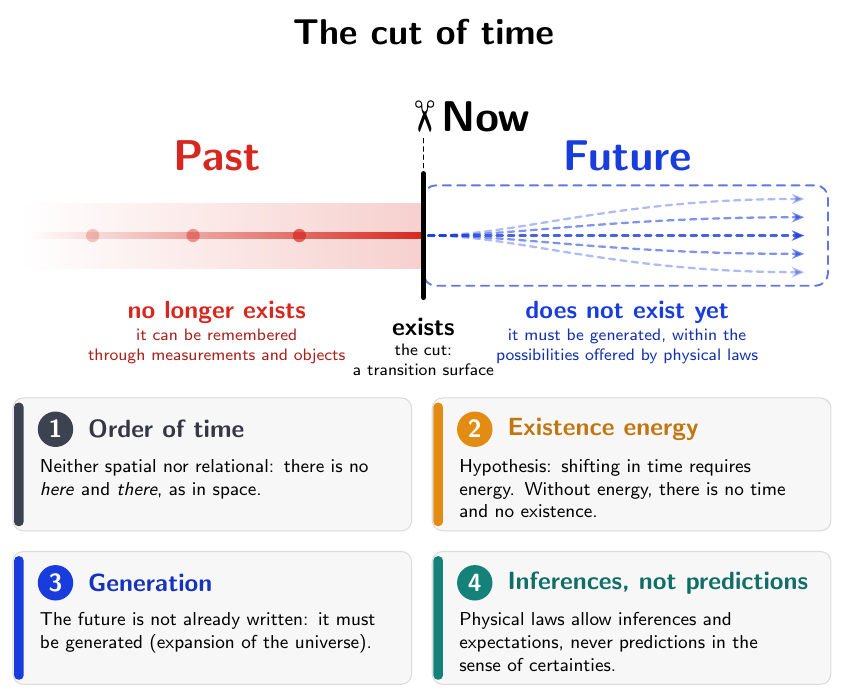}
\caption{The cut of time.}
\label{fig2}
\end{center}
\end{figure}

The order of time is neither spatial, nor relational (differently from the relational view, see e.g. \citealt{ROVELLI}): there is no here, no there, like in space, because the past and the future do not exist as the now. The past has been: one can have memory of it with measurements or with objects, but it does not exist anymore. In astronomy and astrophysics it is common to think that photons from cosmic objects come from the past, so one is doing some kind of archaeology. That is true, but one has to think really as an archaeologist: pyramids were built in the past, but the buildings one can touch today are not exactly the same building of a few thousands of years ago. Matter changed in time. Recently, even the International Prototype of the Kilogram (\emph{Le Grand K}) has been abandoned, because its mass and that of its official copies drifted apart by about 50~$\mu$g in more than one century, and the unit of mass has been redefined via the Planck constant in 2019(\footnote{\url{https://phys.org/news/2019-05-adieu-le-grand-kilogram-redefined.html}.}). For a macroscopic object, it is easy to recognise the decay on long time scales, but it is due to the interactions with the environment (erosion, contamination, chemical reactions), which act in time. It is much more difficult to isolate a possible effect of time alone, without any interaction: for particles, there could be an opportunity (see the next Section).

The future does not exist yet and \emph{must be generated}. This is a key point, a key difference with respect to the block universe or the multiverse. Therefore, the now, the cut of time, is a transition surface open to the infinite possibilities offered by the laws of physics. Now what is possible according to physical laws comes into existence while doing. It is worth noting that, although the future is driven by the laws of physics, it is intrinsically unpredictable, as known from quantum physics and dissipative processes. One can draw inferences by using physical laws, calculate probabilities and expected outcomes, sometimes with an extraordinary accuracy (e.g. the eclipses), but these are never predictions in the etymological sense of certainties, as the future does not exist yet and must be generated. In addition to quantum indeterminacy, chaotic dynamics sets finite horizons to any inference. In this sense, prediction would be just superstition\footnote{To predict derives from Latin \emph{prae-dicere}, to say in advance, to announce with certainty that something in the future will happen. Physicists use the word in a weaker sense, but it would be better to avoid it, unless you are an incurable determinist.}.
A short summary of the idea of time as a cut is displayed in Fig.~\ref{fig2}.

The interpretation proposed in this essay is in partial agreement with Ellis' evolving block universe (e.g. \citealt{ELLIS,ELLIS2,ELLIS3}). The key differences are two: first, Ellis thinks that the past continues to exist, so that there is a block universe continuously growing, made by the past and the now, which in turn is the border surface; second, time is still viewed as duration, not as a cut. Particularly close to the present proposal is the \emph{crystallizing block universe} by \citet{ELLISROTHMAN}, where quantum measurements define the present along a cosmic foliation. Other approaches sharing the idea that time is real and the future is open are those by \citet{SMOLIN}, and by \citet{MULLER}, who links the generation of new time to the expansion of the universe.

\section{The existence energy}\label{sec:exist}
As known from quantum mechanics and relativity, the motion in space is related to the momentum, while that in time is related to the energy. I would like to focus on the time shift only. My hypothesis is that any physical object needs some energy to exist, just to shift in time without any other type of motion or interaction. In other words, a pure time shift would require energy to be done, which would be somehow lost as time goes by. For the moment, I do not care about where this energy goes (some possibilities are discussed in a companion essay, \citealt{NOMASS}). An explicit, although classical, model of a massive object slowly losing its rest energy is the geon, a standing electromagnetic wave held together by its own gravitational attraction, whose trapped radiation leaks away with a calculable lifetime \citep{WHEELER}. 

Existence is just time shift and must not be confused with life. A human being needs a certain amount of calories per day just to live (basal metabolic rate), which is much greater than the energy needed by the particles constituting his body to exist (and that will continue to exist after his death). Lifeless objects also have various types of energy not related to their plain existence. For example, atoms in the lattice of a solid body vibrate because of thermal energy, and also require energy to be bound to each other in the lattice, as well as electrons are bound to atoms. All these particles, regardless their interactions, require a much smaller quantity of energy to simply exist. It is not thermal or chemical or any other type of known energy. I do not know what type of energy could be, and I do not want to go too far with speculations, but there must be some energy to allow a physical body to shift in time without any other motion or interaction. Pure existence. The space too needs some energy to exist, and it could be the vacuum energy (on whether the vacuum itself needs an existence energy, see \citealt{NOMASS}), although, with a cosmological constant, the vacuum energy contained in a comoving volume grows, rather than decreasing, as the universe expands.

It must be noted that momentum and energy are the generators of translations in space and in time, respectively. A generator, however, is not a fuel: a uniform motion in space does not require any energy (inertia) and, by analogy, one could expect that a pure shift in time requires none. Indeed, Noether's theorem states that, if the laws of physics are invariant under time translation, then energy is conserved. Therefore, the hypothesis explored here is equivalent to a tiny violation of the invariance under time translation: the laws of physics would slightly change as time goes by. This is not inconceivable: in an expanding universe there is no global time-translation symmetry, and energy is not globally conserved (e.g. photons lose energy because of the cosmological redshift). In relativity, there is already an energy associated with the pure shift in time: for a body at rest, the four-momentum points entirely in the time direction, and its time component is the rest energy $mc^2$. The hypothesis is then that the rest energy slowly decreases. In this view, mass loses its status of primitive concept, and becomes the energy of fields in their rest frame, a small part of which would be spent to exist at rest \citep{NOMASS}.

One can also try to estimate the energy associated with the expansion of space by an elementary amount. In general relativity there is no global conservation of energy in an expanding universe, so that the answer depends on the accounting, but the local balance for a comoving volume $V$ is well defined: $d(\rho c^2 V) = -p\,dV$, where $\rho$ is the density and $p$ is the pressure. For matter ($p\simeq 0$), the energy contained in a comoving volume does not change: the expansion does not affect the rest energy of particles, so that a possible existence energy would be a new effect, not already included in standard cosmology. For radiation ($p=\rho c^2/3$), the energy decreases because of the cosmological redshift. For a cosmological constant ($p=-\rho c^2$), the energy increases with the volume: $dE = \rho_\Lambda c^2\,dV$. With the measured value $\rho_\Lambda c^2 \simeq 5\times 10^{-10}$~J/m$^3$ \citep{PLANCK2018}, adding one Planck volume $l_{\rm P}^3$ (with $l_{\rm P}\simeq 1.6\times 10^{-35}$~m) to space corresponds to an energy $\rho_\Lambda c^2 l_{\rm P}^3 \sim 2\times 10^{-114}$~J, which is about $10^{-123}$ times the Planck energy: this is the well-known problem of the cosmological constant \citep{WEINBERG}. Within the Hubble volume, about $10^{166}$ Planck volumes are added per second, and the vacuum energy grows by about $4\times 10^{52}$~W. Therefore, with the observed composition of the universe, the expansion does not consume energy, but rather creates vacuum energy. In some formulations, this increase is compensated by a negative gravitational energy, but these are conventions of accounting, not measurements.

This existence energy must be a very small amount, because the matter is stable on long time scales and it was never necessary to include it when applying the principle of energy conservation or any other calculation. I would like to try estimating at least an upper limit. I consider the basic constituents of matter, protons and electrons, without any other type of energy due to interactions. This existence energy should be stored in the rest mass, minus the energies of the products of natural decay. The estimate requires a model. The simplest one is to assume that each particle has a reserve of existence energy $\Delta E$, consumed at a constant power $P$, and that the particle decays when the reserve is exhausted. However, this model implies that particles age: since protons and electrons were created about 13.8 billion years ago, they would all decay at nearly the same moment, and the only constraint is that such a moment has not arrived yet, i.e. $P < \Delta E/t_0$, where $t_0\simeq 4.4\times 10^{17}$~s is the age of the universe. This gives $P\lesssim 3\times 10^{-28}$~J/s for the proton and $P\lesssim 2\times 10^{-31}$~J/s for the electron (with $\Delta E$ as given below). Experimental lower limits on lifetimes, instead, are derived assuming an exponential decay law, which has no memory: a particle that survived for a certain time has the same probability to decay as a newly created one. Dividing $\Delta E$ by these limits, as done below, gives much smaller values, which should be taken only as reference scales, since they combine two different statistical models. The order-of-magnitude calculations for the two cases are as follows:

\begin{enumerate}
\item \emph{Proton:} the decay channel considered dominant in many grand unified theories is $p\rightarrow e^{+}\pi^0$, with a lower limit on the partial lifetime $\tau > 2.4\times 10^{34}$~years \citep{SK2020,RPP}. The difference in energy (mass) between the proton and its decay products (positron and pion) is $\Delta E \sim 1.3\times 10^{-10}$~J. Part of this energy could be redistributed in kinetic energy of particles generated after the vanishing of the proton. Therefore, this $\Delta E$ can be considered an upper limit of the energy available for the existence. The corresponding reference scale for the power needed to shift in time is $\Delta E / \tau \lesssim 1.7 \times 10^{-52}$~J/s.

\item \emph{Electron:} the decay $e^{-}\rightarrow \gamma \nu_{e}$ would violate the conservation of electric charge, so that its search is actually a test of this conservation law; the lower limit of the lifetime is $\tau > 6.6\times 10^{28}$~years \citep{BOREXINO}. The photon would carry about half of the rest energy ($\sim 256$~keV), while the other half would go into the kinetic energy of the neutrino. For consistency with the proton case, $\Delta E$ is the whole rest energy minus the (negligible) rest energy of the neutrino, $\Delta E \sim 8.2\times 10^{-14}$~J, which implies a reference scale for the existence power of $\Delta E / \tau \lesssim 3.9 \times 10^{-50}$~J/s.

\end{enumerate}

With the most stringent limit for the proton, the two values differ by about two orders of magnitude, so there is no coincidence. This is not surprising, since lower limits on lifetimes reflect the exposure of the experiments (number of particles multiplied by the observation time), rather than a property of the particles. Moreover, a uniform decrease of all masses would be unobservable, as it would be equivalent to a change of units: only dimensionless ratios can be measured, such as the proton-to-electron mass ratio $\mu$. The above values imply different fractional rates for the proton and the electron, so that $\mu$ would change by about $10^{-29}$ per year, to be compared with present laboratory limits of the order of $10^{-17}$ per year \citep{LANGE}. Such a small quantity is still far beyond the capability of present-day experimental technology. Nevertheless, the search for the tiniest measurable amount of energy, the quest for the existence energy, could open an interesting research direction, at least for those interested in understanding the physics of time.

\section{Final remarks}\label{sec:final}
In this essay, I have proposed to interpret time as a cut rather than as a duration. The now divides a past that no longer exists from a future that does not exist yet and must be generated, within the possibilities allowed by the laws of physics. Reviewing classical physics, quantum mechanics, relativity, and quantum gravity, I have tried to show that time is neglected or spatialised whenever idealisations are introduced, and that it reappears as soon as dissipation, measurement, matter, or the expansion of the universe are taken into account. In this reading, time reversibility is an idealisation, the cut separates the before and the after of a measurement, the lapse function of a privileged cosmological foliation offers a candidate for the cut in general relativity, and the existence of a physical object may require a tiny amount of energy to shift in time. A complementary perspective, in which mass loses its status of primitive concept in favour of time and energy, is developed in a companion essay \citep{NOMASS}.

It must be stressed that the present work is programmatic. It is mainly an interpretative and philosophical reading of the current situation of physics, rather than a new theory: it does not replace the equations of quantum mechanics or general relativity, and it does not provide an alternative mechanism in place of their formalism. Several points remain open, and some of them have been explicitly indicated in the text: how the generation of entangled particles articulates with the full statistics of quantum mechanics, how a privileged foliation can be reconciled with the gauge freedom of general relativity, whether time is continuous or discrete, and whether the existence energy is anything more than an upper limit.

However, an interpretation is not necessarily sterile. By changing the questions that we ask, it can suggest where to look. Some of the issues discussed here are already open to experiment: the quantum or classical nature of gravity can be tested through gravitationally induced entanglement and the diffusion of space-time; a slow change of the laws in time, such as that implied by an existence energy, can be constrained by comparing atomic clocks over the years; and the asymmetry between past and future in the fundamental laws has already been measured in particle physics. Therefore, I do not exclude that future works may lead to changes at the level of theories and equations as well. In the meaning of the word discussed in Sect.~\ref{sec:qg}, this future is not already written: it has still to be generated.

\section*{Acknowledgements}
I would like to thank dr. Sergio Dalla Val and dr. Anna Spadafora for useful hints on linguistic issues. I also thank Pierluigi Marras for a stimulating exchange on X\footnote{See \url{https://cagliarimonamour.com/2026/11/28/il-coltello-del-tempo/}, in Italian.}, which allowed me to clarify some points.

\end{document}